\documentstyle[12pt,epsfig,fullpage]{article}

\def\be{\begin{equation}}
\def\ee{\end{equation}}
\def\up{\uparrow}
\def\dn{\downarrow}
\def\rt{\rightarrow}
\def\bdn{\Downarrow}
\def\bup{\Uparrow}
\begin{document}
\thispagestyle{empty}

\vskip .6in

{\vbox{\centerline{\large{}}}}
\vskip .3in
\centerline{\bf Polytype Kinetics and Quenching of Spin Chains with}
\centerline{\bf Competing Interactions using Trimer-flip Dynamics}
\centerline{\it Dibyendu Das and Mustansir Barma}
\vskip .3in
\centerline{Department of Theoretical Physics}
\centerline{Tata Institute of Fundamental Research}
\centerline{Homi Bhabha Road, Mumbai 400 005, India.}

\begin{abstract}
We consider the effects of a quench to $T = 0$ on a spin system with axial 
next-next nearest 
neighbour Ising interactions, evolving under a conserved $3$-spin 
flip dynamics. Such a model is motivated by the kinetics of stacking layers in
polytypes near the $3C-6H$ transition. We find that the 
system generically gets arrested in interesting metastable states 
which have inhomogeneously distributed quiescent and active regions.
In such arrested states, the autocorrelation function decays as a stretched
exponential $\sim exp(-(t/\tau_{o})^{1 \over 3})$. The latter feature can 
be understood in terms of a mapping of the dynamics within active stretches
to the well known simple exclusion process of particles on a line, and bounds
can be put on $\tau_o$.    

\end{abstract}

\vskip 0.4in


\vskip 0.1in


\vskip 1.5in

\noindent{Author for Correspondence: Mustansir Barma}

\vskip 0.1in

\noindent{Address: Department of Theoretical Physics, Tata Institute of}

\noindent{Fundamental Research, Homi Bhabha Road, Mumbai 400005, India.}

\vskip 0.1in

\noindent{E-mail: barma@theory.tifr.res.in}

\vskip 0.1in

\noindent{Fax: 091-22-215 2110, Telephone: 2152971}

\newpage

\section{Introduction}

When a statistical system is quenched by suddenly reducing the
temperature $T$, we encounter an interesting situation, namely the
nonequilibrium evolution of a high-$T$ state in a low-$T$ environment. Two
outcomes are possible. The system may approach the equilibrium ground
state, perhaps through a coarsening route. Alternatively, it may get
arrested in a long-lived nonequilibrium state whose characteristics
depend both on the interactions in the system and on gross features of
the dynamics, such as conservation laws. In this paper, we examine the
consequences of quenching a 1-d Ising spin system with competing
interactions, which evolves by a multispin-flip dynamics which conserves
an infinite number of quantities. Both the interactions and the
kinetics are motivated by the use of the extended axial next nearest
neighbour Ising (ANNNI) model to describe polytypism in
close-packed crystals, as explained below.

Within this model, we find that the nature of the arrested state obtained by 
a quench to $T=0$ depends on the degree of competition in the interactions. 
In several cases, the state is one in which there are 
dynamically active and quiescent regions interspersed randomly through
the system. Such inhomogeneously active and quiescent (IQA) states have
interesting dynamical properties, and are the focus of this paper. In
particular, we show that in this state, the autocorrelation function
follows a stretched exponential decay law, and that this can be
related to the dynamics of the simple exclusion process in finite
regions. 

\section{Polytype Kinetics and the ANNNI Model}

Materials such as silicon carbide and zinc sulphide are known to undergo a 
number of transitions from one close-packed arrangement to another when
external conditions such as the temperature and pressure are
changed \cite{verma}. Each crystalline structure is associated with a 
particular sequence of stacking close-packed triangular layers; transitions
between different polytypes correspond to rearrangements of the
stacking sequence. 

Let us denote the three possible arrangements of the triangular lattice 
in a layer by A, B and C \cite{verma}. Stacking of layers are then described 
by sequences such as ABCABC$\cdots$ (3C,
face-centered cubic), ABCACB$\cdots$ (6H, hexagonal close-packed) and 
ABAB$\cdots$ (2H, hexagonal close-packed), each a different
polytype. The polytypic constraint of close-packing implies
that no two successive layers can be of the same type.

The extended ANNNI model with third neighbour interactions 
plays an important role in understanding the occurrence of polytypes 
\cite{yeomans, cheng}. The Hamiltonian is
\be
{\cal H} = -J_1 \sum_i S_i S_{i+1} \ -J_2 \sum_i S_i S_{i+2} \ - J_3
\sum_i S_i S_{i+3} 
\ee
where $J_1, \ J_2$ and $J_3$ are coupling constants, competition
between which can lead to the formation of long-period structures. The
Ising spins $S_i = \pm 1$ in Eq.(1) are related to successive pairs in the
stacking sequence of A's, B's and C's as follows: Pairs AB, BC, CA
correspond to $S_i = 1$, while pairs BA, CB, AC correspond to $S_i =
-1$. Thus for instance the 3C structure is translated to the spin
sequence $\uparrow \uparrow \uparrow \cdots$, the 6H structure is
$\uparrow \uparrow \uparrow \downarrow \downarrow \downarrow \cdots$
while the 2H structure is $\uparrow \downarrow \uparrow \downarrow \cdots$. 
A portion of the ground state phase diagram is shown in
Fig. 1. The phase boundaries shown all correspond to first order
transitions; the heavy lines denote boundaries on which an
exponentially large number of ground state configurations are degenerate
\cite{berreto}.

\begin{figure}[tb]
\begin{center}
\leavevmode
\epsfig{figure=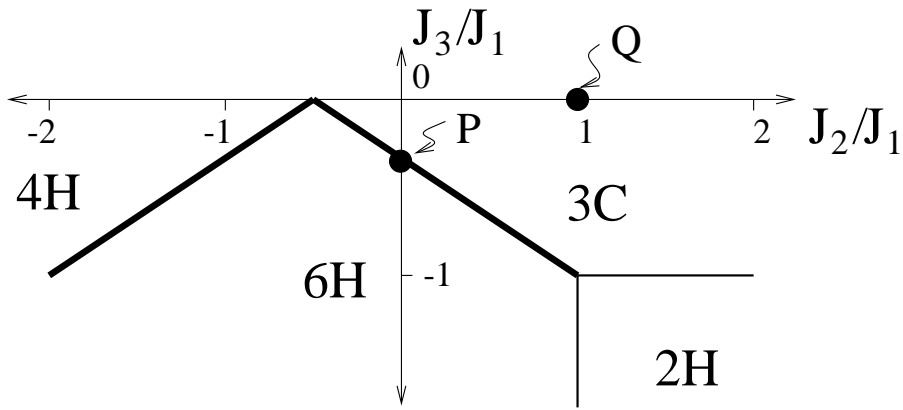,width=12.0cm}
\end{center}
\noindent\caption{A portion of the ground state phase diagram of the third 
neighbour ANNNI model. The lines are first order boundaries; 
thick lines denote multiphase boundaries with an exponentially large number
of degenerate ground states.}
\label{fig:phasedia}
\end{figure}

Besides providing a description of polytype phases, the extended ANNNI
model can also be used to model polytype kinetics. The first such
study was concerned with the kinetic effects of quenching across the
$T=0$ 2H-6H phase boundary \cite{kabra}, and the nature of the resulting 
arrested states. Here we study the effects of quenching the model from $T =
\infty$ to $T = 0$ in the vicinity of the 3C-6H phase boundary. While
modeling kinetics, it is important to allow for only those spin
re-arrangements which correspond to physically admissible moves of the
stacked layers. While local changes in the stacking sequence correspond
to local changes in the spin configuration, the reverse is not
necessarily true. For instance, a single spin flip corresponds to a 
deformation fault which involves the simultaneous movement of a macroscopic
number of layers.

Near the 3C-6H transition, there is experimental evidence \cite{jepps,kabra2}
that the predominant kinetic move is a double-layer displacement 
which involves the simultaneous
movement of two adjacent layers, for example, $\cdots CABC \cdots \rt 
 \cdots CBAC \cdots$ ; two successive layers in the
sequence can interchange provided that such an interchange does not
violate the polytypic constraint of no successive
occurrences of the same letter. In spin language, the
double-layer-displacement move corresponds to flipping triplets of
successive spins, e.g. $\cdots \ \uparrow \uparrow \downarrow
\downarrow \downarrow \uparrow \downarrow \ \rightarrow \ \uparrow
\uparrow \uparrow \uparrow \uparrow \uparrow \downarrow \ \rightarrow
\ \uparrow \downarrow \downarrow \downarrow \uparrow \uparrow \uparrow
\ \cdots$. This is a generalization of the familiar single-spin-flip
Glauber dynamics, but there is an important qualitative difference
between the two. While Glauber dynamics in nonconserving, it turns out
that triplet flip dynamics leaves invariant an infinite number of
conserved quantities \cite{dhar,dhar1}. These conservation laws have important
consequences for time-dependent properties. 

\section{Quenching from $T = \infty$ to $T = 0$ : the IQA State}

Our problem involves studying the effect of quenching to $T = 0$
a system with competing interactions and conservation laws. The
dynamics consists of stochastic flips of triplets, and the $T=0$
environment allows only moves with energy $\Delta E \leq 0$.

\begin{figure}[tb]
\begin{center}
\leavevmode
\epsfig{figure=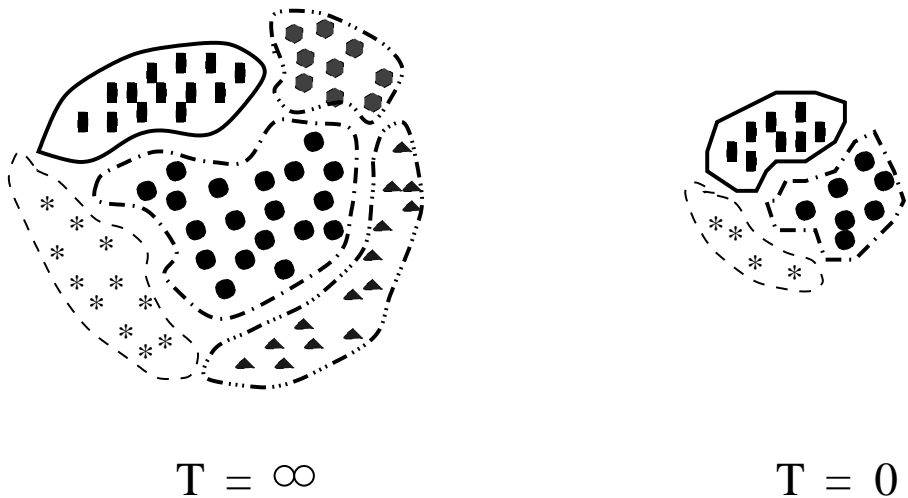,width=12.0cm}
\end{center}
\noindent\caption{Schematic depiction of partitioning of phase space under 
trimer-flip dynamics at $T = \infty$ and $T = 0$. Configurations which can be
reached from each other through the dynamics are shown with the same symbol. 
The set of dynamically connected configurations defines a sector. The 
additional constraints that operate at $T = 0$ lead to a reduction in the number of configurations as well as the number of sectors.}

\label{fig:phasespace}
\end{figure}

Figure 2 is a schematic depiction of the available phase space (i) at
$T = \infty$, and (ii) along the $3C-6H$ multiphase phase boundary 
at $T = 0$, where any
configuration with parallel spin stretches with $3$ or more spins is an
allowed ground state. At $T = \infty$, it is known \cite{dhar,dhar1}
that the infinity of conservation
laws leads to a partitioning of the $2^L$-large phase space into a
large number $N_L$ of sectors; $N_L$ grows as $\sim \mu^L$ where $\mu
\simeq 1.618$ is the golden mean. In case (ii), by contrast, using a transfer 
matrix method we find that the total
number of configurations is reduced to $\sim {\kappa}^L$ with $\kappa \simeq 
1.466$. This ground state subspace is further partitioned into 
${\tilde N}_{L} \sim {\tilde\mu}^L$ sectors, with $\tilde\mu \simeq 1.167$ 
as a consequence of the infinitely many conservation laws.

Evolving the system from a random initial configuration, with the
$\Delta E \leq 0$ condition enforced, corresponds to a rapid quench
from $T = \infty$ to $T = 0$. The energy non-raising condition imposes local
constraints on whether or not a pair of chosen spins can actually be flipped; 
these constraints are a function of $J_{1}, J_{2}$ and $J_{3}$. The normalized
energy changes ${\Delta e} \equiv {\Delta E}/J_1$ involved in flipping a 
triplet $\up\up\up$ to $\dn\dn\dn$ depend on the environments of the triplet,
and are given in Table $1$. There are $20$ distinct local environments; the
other unlisted environments (out of a total of $64$ possibilities) are 
related to these by reflection symmetries.

\begin{center}

\begin{tabular}{|c|c|}
\multicolumn{2}{c}{\bf Table 1: Energy changes for triplet flips in 
various environments} \\[3mm] \hline 
Dynamical moves & $\Delta e $ \\ \hline
$~~\bup\bup\bup\up\up\up\bup\bup\bup$~~$\rt$~~$\bup\bup\bup\dn\dn\dn\bup\bup\bup$~~&~~$4(1 + 2j_{2} + 3j_{3})$~~\\ \hline
$~~\bup\bup\bup\up\up\up\bdn\bdn\bdn$~~$\rt$~~$\bup\bup\bup\dn\dn\dn\bdn\bdn\bdn$~~&~~$0$~~\\ \hline
$~~\bup\bup\bup\up\up\up\bup\bup\bdn$~~$\rt$~~$\bup\bup\bup\dn\dn\dn\bup\bup\bdn$~~&~~$4(1 + 2j_{2} + 2j_{3})$~~\\ \hline
$~~\bup\bup\bup\up\up\up\bup\bdn\bup$~~$\rt$~~$\bup\bup\bup\dn\dn\dn\bup\bdn\bup$~~&~~$4(1 + j_{2} + 2j_{3})$~~\\ \hline
$~~\bup\bup\bup\up\up\up\bup\bdn\bdn$~~$\rt$~~$\bup\bup\bup\dn\dn\dn\bup\bdn\bdn$~~&~~$4(1 + j_{2} + j_{3})$~~\\ \hline
$~~\bup\bup\bup\up\up\up\bdn\bup\bup$~~$\rt$~~$\bup\bup\bup\dn\dn\dn\bdn\bup\bup$~~&~~$4(j_{2} + 2j_{3})$~~\\ \hline
$~~\bup\bup\bup\up\up\up\bdn\bup\bdn$~~$\rt$~~$\bup\bup\bup\dn\dn\dn\bdn\bup\bdn$~~&~~$4(j_{2} + j_{3})$~~\\ \hline
$~~\bup\bup\bup\up\up\up\bdn\bdn\bup$~~$\rt$~~$\bup\bup\bup\dn\dn\dn\bdn\bdn\bup$~~&~~$4j_{3}$~~\\ \hline
$~~\bup\bup\bdn\up\up\up\bup\bup\bdn$~~$\rt$~~$\bup\bup\bdn\dn\dn\dn\bup\bup\bdn$~~&~~$4(j_{2} + j_{3})$~~\\ \hline
$~~\bup\bup\bdn\up\up\up\bup\bdn\bup$~~$\rt$~~$\bup\bup\bdn\dn\dn\dn\bup\bdn\bup$~~&~~$4j_{3}$~~\\ \hline
$~~\bup\bup\bdn\up\up\up\bup\bdn\bdn$~~$\rt$~~$\bup\bup\bdn\dn\dn\dn\bup\bdn\bdn$~~&~~$0$~~\\ \hline
$~~\bup\bup\bdn\up\up\up\bdn\bup\bup$~~$\rt$~~$\bup\bup\bdn\dn\dn\dn\bdn\bup\bup$~~&~~$4(-1 + j_{3})$~~\\ \hline
$~~\bup\bup\bdn\up\up\up\bdn\bup\bdn$~~$\rt$~~$\bup\bup\bdn\dn\dn\dn\bdn\bup\bdn$~~&~~$-4$~~\\ \hline
$~~\bup\bup\bdn\up\up\up\bdn\bdn\bup$~~$\rt$~~$\bup\bup\bdn\dn\dn\dn\bdn\bdn\bup$~~&~~$-4(1 + j_{2})$~~\\ \hline
$~~\bup\bdn\bup\up\up\up\bup\bup\bdn$~~$\rt$~~$\bup\bdn\bup\dn\dn\dn\bup\bup\bdn$~~&~~$4(1 + j_{2} + j_{3})$~~\\ \hline
$~~\bup\bdn\bup\up\up\up\bup\bdn\bup$~~$\rt$~~$\bup\bdn\bup\dn\dn\dn\bup\bdn\bup$~~&~~$4(1 + j_{3})$~~\\ \hline
$~~\bup\bdn\bup\up\up\up\bdn\bup\bdn$~~$\rt$~~$\bup\bdn\bup\dn\dn\dn\bdn\bup\bdn$~~&~~$0$~~\\ \hline
$~~\bup\bdn\bup\up\up\up\bdn\bdn\bup$~~$\rt$~~$\bup\bdn\bup\dn\dn\dn\bdn\bdn\bup$~~&~~$-4j_{2}$~~\\ \hline
$~~\bup\bdn\bdn\up\up\up\bup\bup\bdn$~~$\rt$~~$\bup\bdn\bdn\dn\dn\dn\bup\bup\bdn$~~&~~$0$~~\\ \hline
$~~\bup\bdn\bdn\up\up\up\bdn\bdn\bup$~~$\rt$~~$\bup\bdn\bdn\dn\dn\dn\bdn\bdn\bup$~~&~~$-4(1 + 2j_{2} + j_{3})$~~\\ \hline
\end{tabular}
\end{center}
For all values of $j_2 = J_2/J_1$ and 
$j_3 = J_3/J_1$ for which the $\Delta e$'s listed in Table 1 have the 
same sign, the system will reach an arrested state of the same character.

\begin{figure}[tb]
\begin{center}
\leavevmode
\epsfig{figure=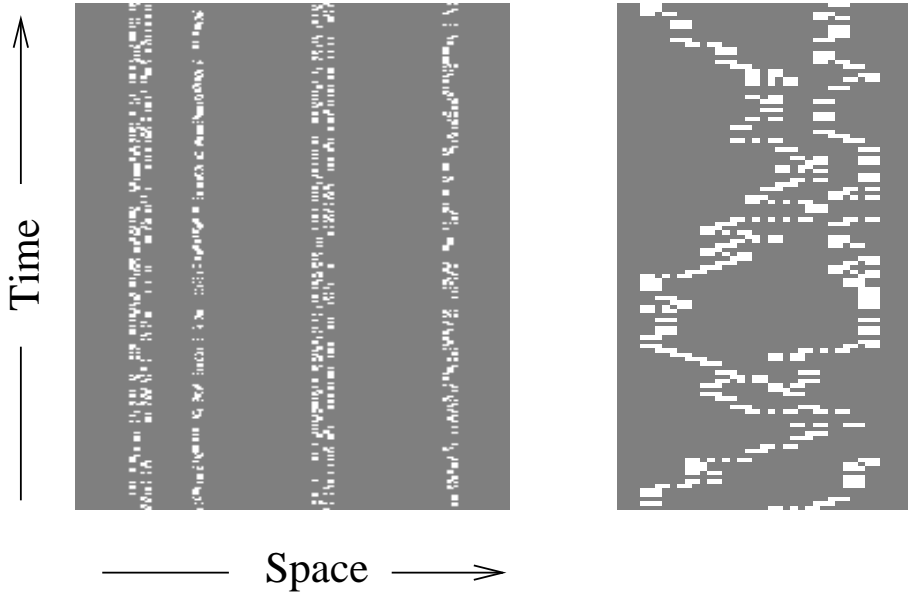,width=12.0cm}
\end{center}
\noindent\caption{The space-time diagram of an IQA arrested state (left) with
activity depicted in white. The activity pattern in a single active stretch
of the IQA state (right) shows the diffusive motion of active walkers, with 
hard core interactions.}
\label{fig:IQA}
\end{figure}

We used numerical
simulations to investigate the properties of arrested states obtained
by quenching to points on the $3C-6H$
boundary, and in the $3C$ phase (points $P$ and $Q$ in Fig. 1, respectively). 
In both cases, the quench does not lead to the ground state. 
For instance, at point $P$, the system does not approach the subspace 
depicted in Fig. 2, but rather an arrested state which is infinitely 
long-lived at $T = 0$. Figure 3 gives an idea of the
nature of this arrested state. The state has
alternating quiescent and active stretches of random lengths. The
lengths of active stretches have an exponential distribution, $A_{P}(\ell)\sim 
 exp (-{\lambda_{P}}\ell)$ with $\lambda_P \simeq 0.22$. We
monitored the decay of the spin autocorrelation function $C_{P}(t)$ in
this state (Fig. 4), and found that it decays as a stretched
exponential $\sim \exp (-(t/\tau_{P})^{1/3})$. We repeated the quench
for a point $Q$ (Fig. 1) in the $3C$ phase, away from the multiphase
line. The equilibrium ground state is $\uparrow \uparrow \uparrow
\uparrow \cdots$ in the case $Q$, in contrast to the large degeneracy of  
ground states at $P$. Nevertheless, on quenching towards $Q$, we found again an
IQA state with an exponentially distributed number of active stretches, 
$A_{Q}(\ell)\sim 
 exp (-{\lambda_{Q}}\ell)$ with $\lambda_{Q} \simeq 0.13$. Further, the 
autocorrelation function $C_{Q}(t)$ decays as $\sim \exp(-(t/\tau_{Q})^{1/3})$,
 but with $\tau_Q > \tau_P$ (Fig. 4). A study of
the nature of the excitations in the active stretches in the two cases
$P$ and $Q$ shows that they are quite different in
character. Despite this, as argued below, they show the common
feature of diffusing excitations confined to randomly distributed
stretches of finite lengths, and this leads to the same form of stretched
exponential decay of $C(t)$ in both cases. 

\begin{figure}[tb]
\begin{center}
\leavevmode
\epsfig{figure=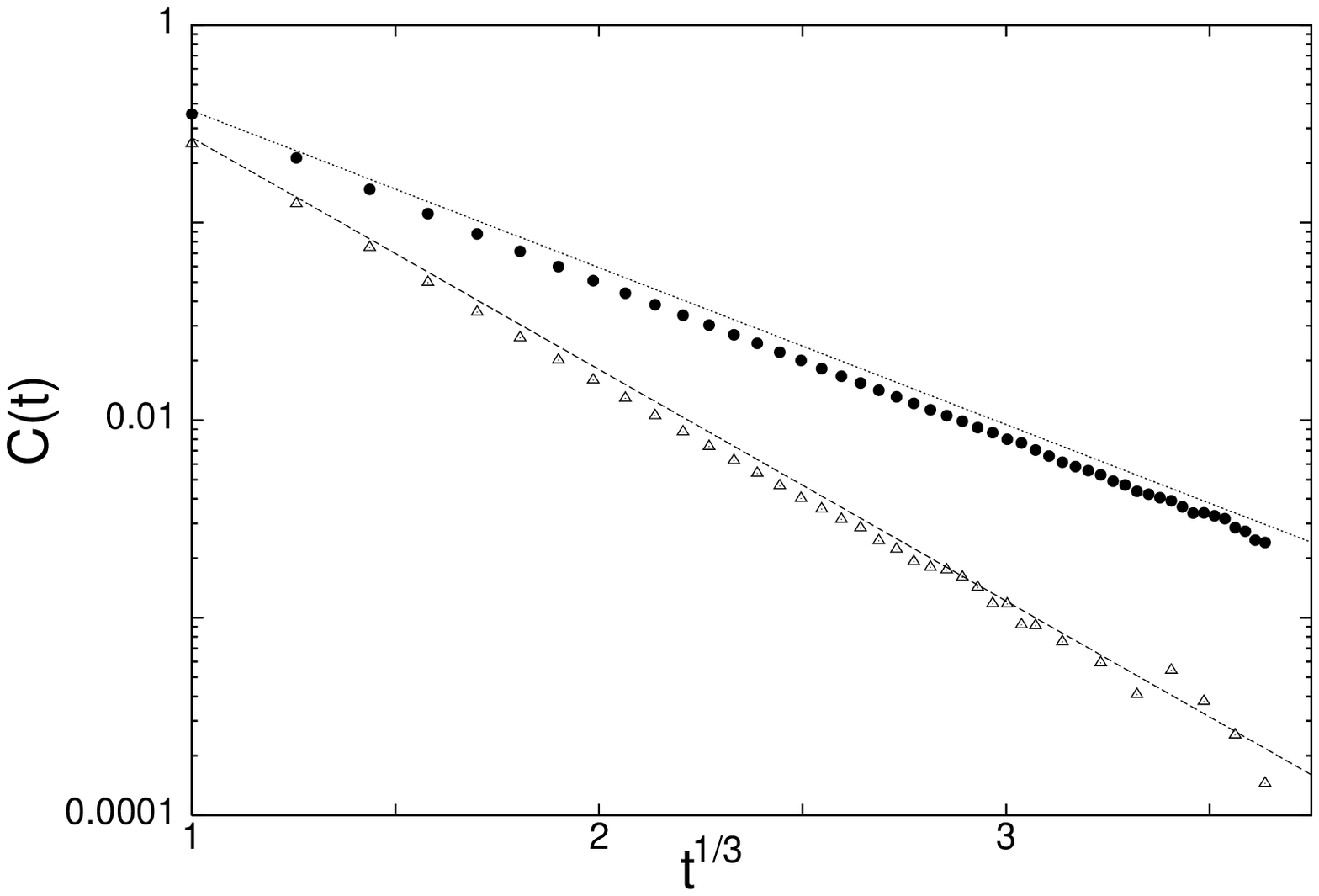,width=16.0cm}
\end{center}
\noindent\caption{Autocorrelation functions $C(t)$ for the two types of 
IQA arrested states at the two points $Q$ (filled circles) and $P$ (empty
triangles) decay as $exp(-(t/\tau_{o}^X)^{1 \over 3})$.}
\label{fig:Corr}
\end{figure}

{\it Quench to $P$} ($j_2 = 0, j_3 = -1/3$, on $3C-6H$ boundary): In this 
case, the active regions consist of parallel spin
patches (PSPs) in a background which is one of two types: (a) antiparallel 
doublets $\up\up\dn\dn\up\up\cdots$ (b) antiparallel doublets with  some 
$\up\dn$ pairs interspersed between doublets, e.g. 
$\up\up\dn\up\dn\dn\up\up\cdots$. 
Flipping a triplet at the edge of a PSP is an
allowed $(\Delta E = 0)$ move, and results in a translation of the PSP
by either one or two lattice spacings, e.g. $\up\up\dn\dn\up\up\up\up\up
\dn\dn \rightarrow \up\up\dn\dn\dn\dn\dn\up\up\dn\dn$. In the process, 
the size of the PSP may fluctuate between 4 and 5, depending on the
surroundings, e.g. $\up\up\dn\up\dn\dn\dn\dn\dn\up\up \rt \up\up\dn\up\up\up\up\dn\dn\up\up$. Backgrounds of type (b) can support only a single PSP, as 
two PSPs around an $\up\dn$ pair lead to a frozen structure. By contrast,
backgrounds of type (a) can support several mobile PSPs. In this case, 
two PSPs have a hard core repulsion and cannot pass through each 
other. These statements can be  easily checked using Table $1$, since 
$\Delta E > 0$ moves are disallowed. Thus we have a 
gas of extended objects with hard core interactions, diffusing in a
confined region. The boundaries of quiescent regions have configurations 
only of a few types --- for example, one such left-boundary is
$\up\up\dn\dn\dn...$, which is stable  
with respect to the adjacent active stretches described above (again 
specific moves can be checked using Table $1$).

{\it Quench to $Q$} ($j_2 = 1, j_3 = 0$, in $3C$ phase): In this case, 
active regions consist of 
stretches of ground state configurations, separated by domain walls, e.g.
$\cdots \up\up\up\up | \dn\dn\dn\dn\dn | \up\up\up\up \cdots$. The spacing 
between successive domain walls is $(3m + 2)$ with $m = 0,1,2,...$. A triplet
adjacent to a wall can flip, whereupon the wall moves by three lattice 
spacings. Two diffusing walls exhibit a short range repulsive interaction, 
as they can approach no closer than two lattice spacings. Once again, the
dynamics is tantamount to a gas of objects with hard core repulsion diffusing
in a confined region. The quiescent boundaries in this case are also of 
a few types, e.g. a stable left-boundary of a quiescent region is 
$\up\up\dn\up\up...$; it can be checked that any attempt to flip its 
edge spins leads to energy raising.

In the active stretches in both cases $P$ and $Q$, we have diffusing walkers 
with hard core repulsion confined to finite regions between impenetrable
walls. Let the number density of such 
walkers in any active stretch be $\tilde{\rho}$. The autocorrelation function  
${C_{\ell}}^X(\tilde\rho,t)$ (where $X$ stands for either $P$ or $Q$), of such 
a stretch is expected to decay at large times as $exp(-t/\tau_{X})$. Here 
$\tau_{X}$ is the relaxation time which depends on the size $\ell$ of the 
finite stretch.
A crucial point to note is that the diffusing walkers are extended objects
--- the PSPs in case $P$ have finite size, and the domain walls in case 
$Q$ have finite range of repulsion between them. In spite of this, in both 
cases   
the dynamics can be mapped exactly to the well known simple exclusion process 
(SEP) \cite{liggett} of particles on a line of a reduced length $\ell_{X} = 
r_X \ell^{'}$. Here $\ell^{'}$ differs slightly from $\ell$, 
and is defined below.

In case $P$, for a background of type (a), every PSP is mapped onto a single
particle, while a doublet is mapped onto a hole $(\up\up\dn\dn\up\up\up\up\up\dn\dn\up\up\up\up\up\dn\dn \rt \circ\circ\bullet\circ\bullet\circ)$ leading to
$r_X = (1 - 3 \rho)/2$. In the case of backgrounds of type (b), 
which have only a single PSP, $\ell_X$ depends on the number of $\up\dn$ pairs 
present; if the maximum number of such pairs are allowed 
(e.g. $\up\up\dn\up\dn\dn\up\dn\up\up...$), $r_X = 3/4$, 
as every doublet and a single spin gets 
mapped to a hole. In case $Q$, every parallel triplet 
is replaced by a hole and an $\up\dn$ pair associated with a domain wall is 
replaced by
a particle ($\up\up\up\up\dn\dn\dn\dn\dn\dn\dn\dn\up\up\up\up \rt \circ\bullet\circ\circ\bullet\circ$) leading to $r_X = (1 + \rho)/3$.  
The modified length $\ell^{'}$  is ($\ell + 4$) and ($\ell + 2$)
in the cases $P$ and $Q$ respecively. The difference between $\ell$ and
$\ell^{'}$ arises from consideration of the dynamics of the 
walkers near the active-quiescent boundaries; since $\ell^{'} \simeq \ell$ 
for large $\ell$, we have $\rho = {\tilde\rho}\ell/\ell^{'} \simeq \tilde\rho$.

For the mapped SEP with free boundary 
conditions, one knows that the relaxation time $= {\ell_X}^2/{\pi^2}D$ 
where $D = 1/2$ is the diffusion constant of the SEP. 
Hence in our problem $\tau_X = 2r_{X}^{2}{{\ell}^{'}}^2/{\pi^2}$.    
The full autocorrelation function $C^{X}(t)$ is then given by averaging 
over the distribution $A(\ell,\rho)$ of stretches of length $\ell$ and 
walker density $\rho$.    
\be
C^{X}(t) = \sum_{\ell,\rho} A(\ell,\rho) {C_{\ell}}^{X}(\rho,t)
\ee
To make further progress, note that the largest and the smallest values of
$\tau_X$ (as $\rho$ is varied from $1/\ell^{'}$ to its maximum possible value)
provide bounds on ${C_{\ell}}^X(\rho,t)$, and thereby on $C^{X}(t)$. Using 
$A(\ell) \sim exp(-\lambda_X \ell)$ and $\tau_X = 
2{r^*}_{X}^{2}{\ell^{'}}^2/{\pi^2}$,
(where ${r^*_X}$ is a different constant for the minimum and maximum 
densities), we see that the above sum is dominated by a saddle point value 
$\ell^* = 
(\pi^2 t/\lambda_X{r^*}_{X}^{2})^{1 \over 3}$. This leads to a stretched 
exponential form for both the upper and lower bounds, implying that
$C^X(t) \sim exp(-(t/\tau^{X}_{o})^{1 \over 3})$. In case $P$, the lower and 
upper bounds on $\tau^{P}_{o}$ are $(1/5)^2({8/(27 \pi^2 \lambda_{P}^{2})})$
(corresponding to $\rho = 1/5$) and ${8/(27 \pi^2 \lambda_{P}^{2})}$ 
(corresponding to $\rho = 0$). In case $Q$, they are $(1/2)^2({8/(27 \pi^2 
\lambda_{Q}^{2})})$ (corresponding to $\rho = 1/2$) and ${8/(27 \pi^2 
\lambda_{Q}^{2})}$ (corresponding to $\rho = 0$). 
Inserting numerical values $\lambda_P = 0.22$ and $\lambda_Q = 0.13$, we
have $0.024 \leq \tau_{o}^{P} \leq 0.61$ and $0.44 \leq \tau_{o}^{Q} \leq 
1.75$. This helps us to
understand the reason for the numerically observed slower decay ($\tau^{P}_{o}
 = 0.37$ and $\tau^{Q}_{o} = 0.56$ in Fig. 4) of $C^Q(t)$ compared to 
$C^P(t)$.

\section{Conclusion}

In our study of $T = \infty \rt T = 0$ quenches of the extended ANNNI model
evolving under triplet flip dynamics, we found interesting arrested states
with interspersed quiescent and active regions. Although the character of 
excitations is quite different in the two cases studied, the autocorrelation 
function shows similar stretched exponential decays. This was traced to the 
fact that active regions in both cases carry diffusing excitations, and  
the dynamics in both cases can be mapped on to the simple exclusion process.

We have found and studied IQA states in quenches of other models including 
the ANNNI model
with Kawasaki dynamics. In this case, in addition to the stretched 
exponential decay of $C(t)$ in steady state, we found an interesting $2$-step
relaxation towards equilibrium when the temperature is raised slightly 
\cite{barma}.

It would be interesting to try to identify signatures of IQA states in rapid
quenches of polytypic materials, for instance near the $3C-6H$ phase 
boundary.

{\bf Acknowledgements}: MB is grateful to Prof. Dhananjai Pandey for providing
an insightful introduction to polytype kinetics and for several discussions
on the subject over the years.

\end{document}